\documentstyle[12pt]{article}

\def\dis{\displaystyle}
\def\del{\partial}
\def\noi{\noindent}
\def\til{\tilde}
\def\wtil{\widetilde}
\def\Sig{\Sigma}

\def\a{\rm a}
\def\b{\rm b}

\begin{document}

\begin{center}
{\Large\bf On the canonical formalism of $f(R)$-type gravity using Lie 
derivatives}\\[7mm]
\end{center}
Yoshiaki Ohkuwa${}^{(1)(*)}$ and Yasuo Ezawa${}^{(2)(**)}$\footnote{${}^{(*)}$ E-mail:
ohkuwa@med.miyazaki-u.ac.jp\\
\hspace*{6mm}${}^{(**)}$ E-mail: ezawa@phys.sci.ehime-u.ac.jp}
\\[5mm]
${}^{(1)}$ Section of Mathematical Science, Dept. of Social Medicine, Faculty 
of Medicine,\\
\hspace*{5mm}University of Miyazaki, Kihara, Kiyotake-cho, Miyazaki, 889-1692, 
Japan\\[1mm]
${}^{(2)}$ Dept. of Physics, Ehime University, Bunkyo-cho, Matsuyama 790-8577, 
Japan\\

\noi
\hspace*{1cm}\begin{minipage}{14.5cm}
{\bf Abstract}\\[1mm]
We present a canonical formalism of the $f(R)$-type gravity using the Lie 
derivatives instead of the time derivatives by refining the formalism of 
our group.
The previous formalism is a direct generalization of the Ostrogradski's 
formalism.
However the use of the Lie derivatives was not sufficient in that Lie 
derivatives and time derivatives are used in a mixed way, so that the 
expressions are somewhat complicated.
In this paper, we use the Lie derivatives and foliation structure of the spacetime 
thoroughly, which makes the procedure and the expressions far more concise.
\\[6mm]
\end{minipage}
\\

\section{Introduction}

The $f(R)$-type gravity is one of the generalized gravities and since
its use by Caroll et al.\cite{CDTT} to explain the discovered accelerated 
expansion of the universe
\cite{Accel1,Accel2,Accel3,Accel4,Accel5,Accel6,
Accel7,Accel8,Accel9,Accel10,Accel11,Accel12}, 
the theory confronted with 
observations and has been attracting much attention and its various 
aspects and applications have been investigated
\cite{S-F, N-O1,N-O2,N-O3,Cap,Que}.
Before that the generalized gravity theories have mainly been interested 
in because of their theoretical advantages.
Main advantageous points are : (i) The theory of gravitons can be 
renormalizable contrary to the Einstein gravity\cite{U-DW,Stelle}. 
(ii) It might be possible to avoid the initial singularity of the universe
\cite{Nariai1, Nariai2} 
proved by Hawking\cite{Haw}. 
(iii) The inflation of the universe at early stage could be explained without 
introducing an ad hoc scalar field\cite{Staro}.

However, the canonical formalism of the $f(R)$-type gravity had not been 
so systematic since it is a somewhat complicated higher derivative theory.
So in the previous paper\cite{EIOWYY}, our group proposed a formalism by 
directly generalizing the canonical formalism of Ostrogradski\cite{Ost} by 
using the Lie derivatives instead of the time derivatives.
The generalization is necessary since the scalar curvature $R$ depends 
on the time derivatives of the lapse function and shift vector.
So, if the Ostrogradski's formalism is directly applied, these variables 
have to obey field equations.
Then only the solutions to these equations are allowed.
This, however, is in conflict with general covariance since a set of 
these variables specify a coordinate frame, so should be taken arbitrarily.
One of the ways to resolve this problem had been given by Buchbinder 
and Lyakhovich(BL method)\cite{B-L1,B-L2} which is sometimes referred to as the 
generalization of the Ostrogradski's one\cite{D-S-Y}.
However, BL method has an undesirable property that, when the generalized 
coordinates are transformed, the Hamiltonian is also transformed
\cite{EIOWYY}.

The formalism in \cite{EIOWYY} remedied the undesirable property of the BL 
method due to the direct generalization of the Ostrogradski's one.
However the Lie derivatives and time derivatives are used in a mixed way, 
so some expressions are complex.
Here all the time derivatives are replaced by the Lie derivatives, which 
makes the procedure more transparent and the resulting expressions far 
more concise. 
We also note that, while the time derivatives depend on the coordinate 
frame so that the differentiated quantities are not tensors generally, 
Lie derivatives preserve the tensorial property, so would be favorable 
derivatives for relativistic treatment.
Furthermore, in ADM formalism\cite{ADM}, which we use here, foliations of 
the $d$-dimensional spacelike hypersurfaces are connected by a one-parameter 
mapping, so that Lie derivatives are natural to express the rate of changes 
and in the definition of the Hamiltonian density.

Finally, we note that surface term, Gibbons-Hawking\cite{surface1,surface2} term, 
which is necessary in Einstein gravity for the variational principle to 
be consistent, is not necessary in the $f(R)$-type gravity\cite{EO}.

In section 2, we review the canonical formalism of Ostrogradski to help seeing 
that our method is a direct generalization of the Ostrogradski' formalism.  
In section 3, we present our formalism. 
In section 4, we show the invariance of our Hamiltonian under the 
transformation of the generalized coordinates. 
Section 5 is devoted to the summary and discussions.

\section{Review of the Ostrogradski's canonical formalism}

In this section, we review the canonical formalism of Ostrogradski which is 
the generalization of the formalism for usual systems to the system described 
by the Lagrangian with higher order time derivatives.\footnote{This section is 
essentially the first half of \cite{EIOWYY} which in turn depends heavily on a 
book by Kimura T. and Sugano R.\cite{K-S}}
We also expect that this section would be helpful in seeing that our formalism 
given in the next section is a direct generalization of the Ostrogradski's one.

Let us consider a system with $N$ degrees of freedom, the generalized 
coordinates of which will be denoted as $q^i,\ (i=1,2,\cdots,N)$.
Its Lagrangian $L$ is assumed to be dependent on these generalized coordinates 
and their time derivatives up to $n$-th order.
The $N(n+1)$ dimensional space, which will be called as velocity phase space, 
has coordinates which are the arguments of the Lagrangian:
$$
D^sq^i\ (s=0,1,\ldots,n;\ i=1,\cdots, N),                                                      \eqno(2.1)
$$
where $\dis D\equiv {d\over dt}$, i.e., the Lagrangian is assumed 
to be defined in the velocity phase space\cite{K-S};
$$
L=L(q^i,\dot{q}^i,\ldots,q^{i(n)})\equiv L(D^sq^i).                                           \eqno(2.2)
$$
It is possible to generalize further that $n$ is different for different 
$i$, but we do not think of this possibility for simplicity.
Transition to the canonical formalism is given by the Ostrogradski 
mapping, which might be seen as the straightforward generalization of 
the Legendre mapping for systems with $n=1$.

Take the following variation of the action $S$:
$$
\delta S\equiv 
\int_{t_{1}+\delta t_{1}}^{t_{2}+\delta t_{2}}L(q^i+\delta q^i,\ldots,
q^{i(n)}+\delta{q}^{i(n)})dt
-\int_{t_{1}}^{t_{2}}L(q^i,\ldots,q^{i(n)})dt                                                          \eqno(2.3)
$$                                                                  
where
$$
\delta q^i\equiv (q+\delta q)^i(t+\delta t)-q^i(t)\approx
\delta^{*}q^i+\dot{q}^i\delta t\ \ \ {\rm with}\ \ \ \delta^{*}q^i
\equiv (q+\delta q)^i(t)-q^i(t),                                                                     \eqno(2.4)
$$
i.e., $\delta^{*}q^i$ is equal to the variation used in the variational 
principle.
Then we have
$$
\delta S=\Bigl[L\delta t\Bigr]_{t_{1}}^{t_{2}}+\delta^{*}S,                                   \eqno(2.5)
$$
where we used the approximation
$$
\int_{t_{k}}^{t_{k}+\delta t_{k}}L(q^i,\ldots, q^{i(n)})dt
=L\left(q^1(t_{k}),\ldots, q^{i(n)}(t_{k})\right)\delta t_{k},
\ \ \ (k=1,2).                                                                                          \eqno(2.6)
$$
In $\delta^{*}S$, the effects of the variation of the time 
$\delta t$ are assumed to be negligible, so  $\delta^{*}S$ reduces 
to the variation taken in the variational principle and is expressed as
$$
\delta^{*}S=\int_{t_{1}}^{t_{2}}\delta^{*}L\,dt
=\Bigl[\delta F\Bigr]_{t_{1}}^{t_{2}}+\til{\delta}S.                                            \eqno(2.7)
$$
By arranging the sum in $\delta^{*}S$, we have
$$
\delta F=\sum_{i=1}^N\Biggl[\,\sum_{s=0}^{n-1}\Biggl\{\sum_{r=s+1}^n
(-1)^{r-s-1}D^{r-s-1}\Bigl({\del L\over \del(D^rq^i)}\Bigr)\Biggr\}
\delta^{*} q^{i(s)}\,\Biggr]                                                                       \eqno(2.8)
$$
and
$$
\til{\delta}S
=\sum_{i=1}^N\int_{t_{1}}^{t_{2}}\sum_{s=0}^{n}(-1)^sD^s\Bigl(
{\del L\over \del(D^sq^i)}\Bigr)\delta^{*}q^idt.                                            \eqno(2.9)
$$
Thus we have
$$
\delta S=\Bigl[L\delta t+\delta F\Bigr]_{t_{1}}^{t_{2}}
+\til{\delta}S,                                                                                       \eqno(2.10)
$$
where in the first term on the right hand side, $\delta^{*}q^i$ is 
written by $\delta q^i$ and $\delta t$ given in (2.4).
The generalized coordinates of the phase space (often referred to as 
the new generalized coordinates for $s\geq 1$) $q^i_{s}$ are 
taken as
$$
q^i_{s}\equiv D^sq^i,\ \ \ (i=1,\cdots,N;s=0,\cdots,n-1),                                \eqno(2.11)
$$
and the momenta canonically conjugate to these coordinates, $p^s_{i}$, 
are defined to be the coefficients of $\delta D^sq^i=\delta q^i_{s}$ 
in $\delta F$:
$$
p^s_{i}\equiv \sum_{r=s+1}^n\left[(-1)^{r-s-1}D^{r-s-1}\left(
{\del L\over \del D^rq^i}\right)\right].                                                     \eqno(2.12)
$$

The Hamiltonian, $H$, is defined as $(-1)\times$(the coefficient of 
$\delta t$) in $L\delta t+\delta F$:
$$
H=\sum_{i=1}^N\sum_{s=0}^{n-1}p^s_{i}Dq^i_{s}-L.                                       \eqno(2.13)
$$
Note that for $s=n-1$, eq. (2.12) has simple expressions:
$$
p^{n-1}_{i}={\del L\over \del \dot{q}^i_{n-1}}.                                 \eqno(2.14)
$$
Thus, it is easily seen that the Ostrogradski mapping 
$$
(q^i,\dot{q}^i,\cdots,q^{i(n)};L)\rightarrow 
(q^i,\cdots,q^{i(n-1)};p^0_{i},\cdots,p_{i}^{n-1};H)                            \eqno(2.15)
$$
is a generalization of the Legendre mapping.
It is noted that the Hamiltonian is invariant under the transformation 
of generalized coordinates ; $q^i\rightarrow q^{'i}$.

\section{Canonical formalism of $f(R)$-type gravity}

In this section, we present a 
canonical formalism of $(1+d)$-dimensional $f(R)$-type gravity by 
refining the formalism proposed in \cite{EIOWYY} which is a direct 
generalization of Ostrogradski's formalism.
As the variables for gravity, we adopt the ADM variables\cite{ADM}, i.e., 
the metric $h_{ij}$ of the $d$-dimensional hypersurface $\Sigma_{t}$ 
which has the normal vector field $n^{\mu}=N^{-1}(1,-N^i)$, $N$,  the 
lapse function and $N^i$, the shift vector.
That is, we regard the spacetime to have the foliation structure\cite{Isham}.
Then the scalar curvature $R$ is expressed in terms of these variables as
$$
R
=h^{ij}{\cal L}_{n}^{\;2}h_{ij}
+{1\over4}\left(h^{ij}{\cal L}_{n}h_{ij}\right)^2
-{3\over4}h^{ik}h^{jl}{\cal L}_{n}h_{ij}{\cal L}_{n}h_{kl}
+{}^d\!R-2\Delta(\ln N).                                                        \eqno(3.1)
$$
${\cal L}_{n}$ represents the Lie derivative along the normal vector 
field $n^{\mu}$. ${}^dR$ is the scalar curvature of $\Sig_{t}$. 
It is noted that $R$ does not depend on ${\cal L}_{n}N$ and 
${\cal L}_{n}N^i$.

\subsection{Variation of the action of the $f(R)$-type gravity}

Lagrangian density of the $f(R)$-type gravity, ${\cal L}_{G}$, is 
taken as
$$
{\cal L}_{G}=\sqrt{-g}f(R),                                                     \eqno(3.2)
$$
where $g\equiv \det{g_{\mu\nu}}$. Then the Lagrangian $L_{G}(\Sig_{t})$ 
and the action $S$ are expressed as follows:
$$
L_{G}(\Sig_{t})=\int_{\Sig_{t}}{\cal L}_{G}\,d^dx,\ \ \ \ \ 
S=\int_{t_{1}}^{t_{2}}L_{G}(\Sig_{t})\,dt
=\int_{t_{1}}^{t_{2}}dt\int_{\Sig_{t}}\,d^dx{\cal L}_{G}.                       \eqno(3.3)
$$
From (3.1) and (3.3), ${\cal L}_{G}$ depends on the ADM variables in 
the following way:
$$
{\cal L}_{G}
={\cal L}_{G}(N,h_{ij},{\cal L}_{n}h_{ij},{\cal L}_{n}^{\;2}h_{ij}).   
                                                                                \eqno(3.4)
$$
It is noted that ${\cal L}_{n}h_{ij}$ is related to the extrinsic 
curvature $K_{ij}$ of $\Sig_{t}$ as
$$
K_{ij}
={1\over2}{\cal L}_{n}h_{ij}
={1\over2N}(\del_{0}h_{ij}-N_{i;j}-N_{j;i}),                                    \eqno(3.5)
$$
where a semicolon denotes the covariant derivative with respect to 
the metric $h_{ij}$.

Now we consider the following variation of the action
$$
\begin{array}{ll}
\delta S&\equiv \dis
\int_{t_{1}+\delta t_{1}}^{t_{2}+\delta t_{2}}dt
\int_{\Sig_{t}}\,d^dx({\cal L}_{G}+\delta{\cal L}_{G})
-\int_{t_{1}}^{t_{2}}dt\int_{\Sig_{t}}\,d^dx\,{\cal L}_{G}
\\[5mm]
&\dis 
=\int_{t_{2}}^{t_{2}+\delta t_{2}}dt\int_{\Sig_{t}}d^dx\,{\cal L}_{G}
-\int_{t_{1}}^{t_{1}+\delta t_{1}}dt\int_{\Sig_{t}}d^dx\,{\cal L}_{G}
+\int_{t_{1}}^{t_{2}}dt\int_{\Sig_{t}}d^dx\,\delta{\cal L}_{G}
\end{array}                                                                     \eqno(3.6)
$$
where
$$
\delta{\cal L}_{G}\equiv 
{\delta{\cal L}_{G}\over\delta N}\delta N
+{\delta{\cal L}_{G}\over\delta h_{ij}}\delta h_{ij}
+{\del{\cal L}_{G}\over\del({\cal L}_{n}h_{ij})}\delta({\cal L}_{n}h_{ij})
+{\del{\cal L}_{G}\over\del({\cal L}_{n}^{\;2}h_{ij})}
\delta({\cal L}_{n}^{\;2}h_{ij}).                                               \eqno(3.7)
$$
Here the first two terms on the right-hand side are not the partial 
derivatives but the functional derivatives since the scalar curvature 
$R$ depends on the derivatives of $N$ and $h_{ij}$ in $\Delta(\ln N)$ 
and ${}^dR$ as seen in (3.1).
Since the correspondence of points on each $\Sig_{t}$ are made by 
one-parameter transformation, derivative with respect to $t$ is given 
by the Lie derivatives along the timelike curves, ${\cal L}_{t}$ e.g.,:
$$
h_{ij}({\bf x},t+dt)=h_{ij}({\bf x},t)+{\cal L}_{t}h_{ij}\,dt,                  \eqno(3.8)
$$
to the first order in $dt$.
So variations of gravitational variables are as follows:
$$
\left\{\begin{array}{l}
\delta h_{ij}\equiv (h_{ij}+\delta h_{ij})(t+\delta t)-h_{ij}(t)
=\delta^{*}h_{ij}(t)+{\cal L}_{t}h_{ij}(t)\,\delta t
\\[3mm]
\delta N\equiv (N+\delta N)(t+\delta t)-N(t)
=\delta^{*}N(t)+{\cal L}_{t}N(t)\,\delta t
\end{array}\right.                                                              \eqno(3.9)
$$
where we note that $\delta^{*}$ variations are those given in (2.4). 
When we use (3.7) in (3.6), 
``partial integrationsh have to be done for terms including 
${\cal L}_{n}\delta h_{ij}$ and ${\cal L}_{n}^{\;2}\delta h_{ij}$.
This is done by using a relation for a scalar field $\Phi$ and 
tensor fields $T^{ij}$ and $S_{ij}$:
$$
\left\{\begin{array}{l}
{\cal L}_{n}(\sqrt{h}N\Phi)
={\cal L}_{n}(\sqrt{h}N)\,\Phi+\sqrt{h}\,N{\cal L}_{n}\Phi
=\del_{\mu}(n^{\mu}\sqrt{h}N\Phi),
\\[3mm]
{\cal L}_{n}(\sqrt{h}NT^{ij}S_{ij})
=\del_{\mu}(n^{\mu}\sqrt{h}NT^{ij}S_{ij}).
\end{array}\right.                                                              \eqno(3.10)
$$
Then we have
$$
\begin{array}{lcl}
\delta{\cal L}_{G}&=&\dis {\delta{\cal L}_{G}\over \delta N}\delta N
+\left[{\delta{\cal L}_{G}\over \delta h_{ij}}-{\cal L}_{n}\Bigl(
{\del{\cal L}_{G}\over \del({\cal L}_{n}h_{ij})}\Bigr)
+{\cal L}_{n}^2\Bigl({\del{\cal L}_{G}\over \del({\cal L}_{n}^2h_{ij})}
\Bigr)\right]\delta h_{ij}
\\[5mm]
&&\dis +\del_{\mu}\left[n^{\mu}\Bigl\{\Bigl({\del{\cal L}_{G}\over 
\del({\cal L}_{n}h_{ij})}-{\cal L}_{n}\Bigl({\del{\cal L}_{G}\over 
\del({\cal L}_{n}^2h_{ij})}\Bigr)\delta h_{ij}+{\del{\cal L}_{G}\over 
\del({\cal L}_{n}^2h_{ij})}\delta({\cal L}_{n}h_{ij})\Bigr\}\right] . 
\end{array}                                                                     \eqno(3.11)
$$
On the right hand side of the second line of (3.6), the first two 
terms are approximated as
$$
\Bigl[L_{G}(\Sig_{t})\,\delta t\Bigr]_{t_{1}}^{t_{2}},                          \eqno(3.12)
$$
and in the last term, it is assumed that effects of the variation of 
the time $\delta t$ can be neglected as in the Ostrogradski's formalism, 
so the $\delta$-variations can be replaced by the $\delta^{*}$ variations.
The situations are the same for eq. (3.11).
Thus the variation of the action $\delta S$ is expressed as
$$
\begin{array}{lcl}
\dis \delta S
&=&\Bigl[L_{G}(\Sig_{t})\delta t\Bigr]_{t_{1}}^{t_{2}}+\til{\delta}S
\\[5mm]
&&\dis +\int_{\Sig_{t}}d^dx\Bigl[\,n^0\Bigl\{{\del{\cal L}_{G}\over 
\del({\cal L}_{n}h_{ij})}
-{\cal L}_{n}\Bigl({\del{\cal L}_{G}\over \del({\cal L}_{n}^2h_{ij})}
\Bigr)\Bigr\}(\delta h_{ij}-{\cal L}_{t}h_{ij}\,\delta t)
\\[5mm]
&&\dis \hspace*{2cm}+n^0\,{\del{\cal L}_{G}\over \del({\cal L}_{n}^2h_{ij})}
\Bigl\{\delta({\cal L}_{n}h_{ij})-{\cal L}_{t}({\cal L}_{n}h_{ij})
\delta t\Bigr\}\,\Bigr]_{t_{1}}^{t_{2}},
\end{array}                                                                     \eqno(3.13)
$$
where
$$
\til{\delta}S
=\int_{t_{1}}^{t_{2}}dt\int_{\Sig_{t}}d^dx\Bigl[\,{\delta{\cal L}_{G}
\over \delta N}\delta^{*}N+\Bigl\{{\delta{\cal L}_{G}\over \delta h_{ij}}
-{\cal L}_{n}\Bigl({\del{\cal L}_{G}\over \del({\cal L}_{n}h_{ij}}\Bigr)
+{\cal L}_{n}^2\Bigl({\del{\cal L}_{G}\over \del({\cal L}_{n}^2h_{ij})}
\Bigr)\Bigr\}\delta^{*}h_{ij}\,\Bigr].                                          \eqno(3.14)
$$

\subsection{Ostrogradski mapping}

Ostrogradski mapping is defined by defining (i) the generalized 
coordinates of the phase space, (ii) momenta canonically conjugate 
to them which are, of course, the coordinates of the phase space 
and the (iii) Hamiltonian density.
These are defined as follows:

\subsubsection{Generalized coordinates}

As the generalized coordinates of the phase space, we take the components 
of the $d$-dimensional metric, $h_{ij}$, which will be referred to as 
original generalized coordinates and those corresponding to $D^{(1)}q^i$, 
as in \cite{B-L1,B-L2}, (a half of) the Lie derivatives of the original 
generalized coordinates which is equal to the extrinsic curvature $K_{ij}$, 
eq. (3.5), and will be referred to as new generalized coordinates and will 
be denoted by $Q_{ij}$.

\subsubsection{Conjugate momenta}

Momenta canonically conjugate to the original and new generalized 
coordinates, denoted as $p^{ij}$ and $P^{ij}$ respectively, are defined 
to be the coefficient of their variations in the total time derivative 
terms in (3.13) and are expressed as follows:
$$
p^{ij}
=n^0\Bigl[{\del{\cal L}_{G}\over \del({\cal L}_{n}h_{ij})}
-{\cal L}_{n}\Bigl({\del{\cal L}_{G}\over \del({\cal L}_{n}^2h_{ij})}
\Bigr)\Bigr],\ \ \ 
P^{ij}=2n^0{\del{\cal L}_{G}\over \del({\cal L}_{n}^2h_{ij})}.                  \eqno(3.15)
$$
Using (3.1) and (3.2), we have the following concrete expressions
$$
\left\{\begin{array}{l}
p^{ij}=-\sqrt{h}\left[{\cal L}_{n}f'(R)h^{ij}+f'(R)Q^{ij}\right],
\\[3mm]
P^{ij}=2\sqrt{h}f'(R)h^{ij},
\end{array}\right.                                                              \eqno(3.16)
$$
where, of course, ${\cal L}_{n}f'(R)$ is also expressed as 
$f''(R){\cal L}_{n}R$.
Decomposing a tensor $T^{ij}$ into traceless and trace parts, 
$T^{\dagger ij}$ and $T$ respectively, as $\dis T^{ij}=T^{\dagger ij}
+{1\over d}h^{ij}T$, we see from (3.16) that $P^{ij}$ has only the 
trace part $P\equiv h_{ij}P^{ij}={2d\sqrt{h}}f'(R)$. 
Solving for $R$, we have
$$
R=f^{'-1}(P/2d\sqrt{h})\equiv \psi(P/2d\sqrt{h}).                               \eqno(3.17)
$$
In other words, we have constraints $P^{\dagger ij}=0$, so only new 
independent canonical pair is 
$$
(Q,{P\over d})\equiv (Q,\Pi),\ \ \ i.e.\ \ \ \Pi\equiv {P\over d} \ .           \eqno(3.18)
$$.

\subsubsection{Hamiltonian density}

Hamiltonian density, ${\cal H}_{G}({\bf x},t)$, is defined as the 
coefficient of $(-1)\times\delta t$ in the time boundary terms of (3.13):
$$
{\cal H}_{G}
=p^{ij}{\cal L}_{t}h_{ij}+P^{ij}{\cal L}_{t}Q_{ij}-{\cal L}_{G}.                 \eqno(3.19)
$$
Concrete expression is also given from eqs. (3.1) and (3.2) as follows:
$$
{\cal H}_{G}=N{\cal H}_{0}+N^i{\cal H}_{i}+{\rm total\ derivatives},               \eqno(3.20)
$$
where, after a canonical transformation $(Q,\Pi)\rightarrow 
(\bar{Q},\bar{\Pi})\equiv (\Pi,-Q)$, ${\cal H}_{0}$ and ${\cal H}_{i}$ 
are expressed as\cite{EO}
$$
\left\{\begin{array}{lcl}
{\cal H}_{0}
&=&\dis {2\over \bar{Q}}\Bigl(p^{ij}p_{ij}-{1\over d}p^2\Bigr)-{2\over d}p\bar{\Pi}
+{1\over2}\bar{Q}\psi(\bar{Q}/2\sqrt{h})-{d-3\over 2d}\bar{Q}\bar{\Pi}^2
-{1\over2}\,{}^d\!R\bar{Q}
\\[5mm]
&&+\Delta\bar{Q}-\sqrt{h}f\left(\psi(\bar{Q}/2\sqrt{h})\right)
\\[5mm]
{\cal H}_{i}&=&\dis 2\Bigl(p^{\ \ ;j}_{ij}-{2\over d}p_{;i}\Bigr)
-\bar{Q}\bar{\Pi}_{;i}+{2\over d}(\bar{Q}\bar{\Pi})_{;i} \ .
\end{array}\right.                                                               \eqno(3.21)
$$
The total derivatives appear when we make partial integration 
in doing the process of variation as usual.

\subsubsection{Canonical equations of motion}

The canonical equations of motion derived from (3.19) are expressed as\cite{Wald}
$$
{\cal L}_{t}h_{ij}={\delta{\cal H}_{G}\over \delta p^{ij}},\ \ \ 
{\cal L}_{t}p^{ij}=-{\delta{\cal H}_{G}\over \delta h_{ij}},                    \eqno(3.22\a)
$$
and
$$
{\cal L}_{t}Q_{ij}={\delta{\cal H}_{G}\over \delta P^{ij}},\ \ \ 
{\cal L}_{t}P^{ij}=-{\delta{\cal H}_{G}\over \delta Q_{ij}}.                    \eqno(3.22\b)
\footnotemark 
$$
\footnotetext{Precisely, eqs. (3.22b) are derived by using the Lagrange multiplier 
method, since they include the constraint equations.}
Since the f(R)-type gravity is massive, or, the graviton has more than two 
polarizations\cite{Pol1, Pol2, Pol3}, we would not modify (3.22a).
However, (3.22b) contain dependent variables, we would extract from them the 
equations for independent variables (3.18). 
Since
$$
P^{ij}{\cal L}_{t}Q_{ij}=\Pi{\cal L}_{t}Q+(-2p^{\dagger ij}
+{1\over d}Q\Pi h^{ij}){\cal L}_{t}h_{ij},                                       \eqno(3.23)
$$
we have
$$
{\cal L}_{G}=p^{ij}{\cal L}_{t}h_{ij}+\Pi{\cal L}_{t}Q-\wtil{\cal H}_{G},        \eqno(3.24)
$$
where
$$
\begin{array}{lcl}
\wtil{\cal H}_{G}&\equiv&\dis {\cal H}_{G}-N\Bigl[{8\over \bar{Q}}\Bigl(p^{ij}p_{ij}
-{1\over d}\,p^2\Bigr)+{2\over d}\bar{Q}\bar{\Pi}^2\Bigr]-N^i\Bigl[4p_{ij}^{\ \ ;j}
-{4\over d}p_{;i}+{2\over d}(\bar{Q}\bar{\Pi})_{;i}\Bigr]
\\[5mm]
&&+{\rm divergent\ terms}.
\end{array}                                                                     \eqno(3.25)
$$
Then the equations for independent variables in (3.22\b) are as follows:
$$
{\cal L}_{t}Q={\delta\wtil{\cal H}_{G}\over \delta\Pi},\ \ \ 
{\cal L}_{t}\Pi=-{\delta\wtil{\cal H}_{G}\over \delta Q}.                    \eqno(3.26)
$$
We can obtain explicit expressions of these equations by using (3.21).
In addition, noting $\wtil{\cal H}_{G}$ as $\wtil{H}_{G}=N\wtil{H}_{0}
+N^i\wtil{H}_{i}$+divergent terms as (3.21), we have
$$
\left\{\begin{array}{lcl}
\dis \wtil{H}_{0}&=&\dis -{6\over \bar{Q}}\Bigl(p^{ij}p_{ij}-{1\over d}\,p^2\Bigr)
-{1\over d}\Bigl(2p\bar{\Pi}+{d+1\over 2}\bar{Q}\bar{\Pi}^2\Bigr)
+{1\over 2}\bar{Q}\psi(\bar{Q}/2\sqrt{h})-{1\over 2}\,{}^d\!R\bar{Q}
\\[4mm]
&&+\Delta\bar{Q}-\sqrt{h}f(\psi(\bar{Q}/2\sqrt{h}))
\\[5mm]
\dis \wtil{H}_{i}&=&-\Bigl(2p_{ij}^{\ \ ;j}+\bar{Q}\bar{\Pi}_{;i}\Bigr) . 
\end{array}\right.                                                                 \eqno(3.27)
$$

\section{Invariance of the Hamiltonian}

We consider the following transformations of the generalized 
coordinates $h_{ij}$:
$$
h_{ij}\rightarrow \phi_{ij}\equiv F_{ij}(h_{kl})\ \ \ 
{\rm or\ inversely}\ \ \ h_{ij}\equiv G_{ij}(\phi_{kl}),                      \eqno(4.1)
$$
and show that the Hamiltonian is invariant under this transformation 
as in the case of Ostrogradski formalism.
New generalized coordinates $\Phi_{ij}$ are defined as in (3.5), i.e.,
$$
\Phi_{ij}\equiv {1\over2}{\cal L}_{n}\phi_{ij}.                                 \eqno(4.2)
$$
Hamiltonian density $\bar{\cal H}_{G}$ expressed in the transformed 
variables is defined to be
$$
\bar{\cal H}_{G}\equiv 
\pi^{ij}{\cal L}_{t}\phi_{ij}+\Pi^{ij}{\cal L}_{t}\Phi_{ij}
-\bar{\cal L}_{G}
(N,\phi_{ij},{\cal L}_{n}\phi_{ij},{\cal L}_{n}^{\;2}\phi_{ij}),                  \eqno(4.3)
$$
where $\pi^{ij}$ and $\Pi^{ij}$ are momenta canonically conjugate 
to $\phi_{ij}$ and $\Phi_{ij}$, respectively. Since
$$
{\cal L}_{n}h_{ij}
={\del G_{ij}\over \del\phi_{kl}}{\cal L}_{n}\phi_{kl},\ \ \ 
{\cal L}_{n}^{\;2}h_{ij}
={\cal L}_{n}\Bigl({\del G_{ij}\over \del\phi_{kl}}\Bigr)
{\cal L}_{n}\phi_{kl}+{\del G_{ij}\over \del\phi_{kl}}
{\cal L}_{n}^{\;2}\phi_{kl},                                                        \eqno(4.4)
$$
$\bar{{\cal L}_{G}}$ is defined as
$$
\bar{\cal L}_{G}(N,\phi_{ij},{\cal L}_{n}\phi_{ij},{\cal L}_{n}^{\;2}h_{ij})
\equiv {\cal L}_{G}\left(N,G_{ij}(\phi_{kl}),{\del G_{ij}\over 
\del\phi_{kl}}{\cal L}_{n}\phi_{kl},{\cal L}_{n}\Bigl({\del G_{ij}\over 
\del\phi_{kl}}\Bigr){\cal L}_{n}\phi_{kl}+{\del G_{ij}\over 
\del\phi_{kl}}{\cal L}_{n}^{\;2}\phi_{kl}\right).                               \eqno(4.5)
$$
$\pi^{ij}$ and $\Pi^{ij}$ satisfy relations similar to (3.15), 
and from these relations, we have
$$
\pi^{ij}=p^{kl}{\del G_{kl}\over \del\phi_{ij}},\ \ \ 
\Pi^{ij}=P^{kl}{\del G_{kl}\over \del\phi_{ij}},                                \eqno(4.6\a)
$$
or inversely
$$
p^{ij}=\pi^{kl}{\del F_{kl}\over \del h_{ij}},\ \ \ 
P^{ij}=\Pi^{kl}{\del F_{kl}\over \del h_{ij}}.                                   \eqno(4.6\b)
$$
With help of (4.6a,b), we have
$$
p^{ij}{\cal L}_{t}h_{ij}
=\pi^{kl}{\del F_{kl}\over \del h_{ij}}{\del G_{ij}\over 
\del\phi_{mn}}{\cal L}_{t}\phi_{mn}=\pi^{ij}{\cal L}_{t}\phi_{ij}.           \eqno(4.7)
$$
Similar relation holds between $P^{ij}$ and $\Pi^{ij}$, so we have
$$
{\cal H}_{G}=\bar{\cal H}_{G}.                                                   \eqno(4.8)
$$
It is noted that the transformation (4.1) includes the coordinate 
transformation on $\Sig_{t}$.

\section{Summary and discussions}

We presented a canonical formalism of the $f(R)$-type gravity by 
generalizing the Ostrogradski's formalism.
Present formalism refines the previous one by our group which remedied 
the undesirable points of BL method, i.e., the Hamiltonian is not 
invariant under the transformation of the generalized coordinates, 
here the metric $h_{ij}$ of the hypersurface $\Sig_{t}$.
In addition, we derived canonical equations of motion for independent 
variables $Q$ and $\Pi$.

The formalism has an important application to the problem of the 
equivalence theorem between the $f(R)$-type gravity and Einstein 
gravity coupled to a scalar field, a kind of the scalar-tensor gravity 
theories. 
To prove the theorem, a conformal transformation depending on the curvature 
is used. 
Therefore the transformation is not necessarily restricted to the canonical 
one in the phase space. 
It was in fact shown that it is not the canonical transformation in 
that the fundamental Poisson brackets before and after the transformation are 
not consistent although the explicit expressions after the transformation are 
not uniquely determined by the original ones\cite{Equiv}.
Thus if the $f(R)$-type gravity were quantized canonically, transformed 
theory should be quantized noncanonically.
This coincides with the result that noncanonical quantization would stabilize 
the extra-dimensional space\cite{ES,ESWY,KJS1,KJS2}.

Present formalism would also be helpful in the investigation of quantum 
gravity.
One version of the theory is that of gravitons which are thought of as 
the duality partner of the gravitational waves which obey the usual wave 
equation.
So it is reasonable that the theory is the canonical quantum theory.
The other version is the quantum cosmology which is the canonical quantum 
mechanics of spacetime and the basic equation is the Wheeler-DeWitt(WDW) 
equation.
However, no dual partner of spacetime is known definitely.
Finally, in order to avoid the probabilistic problem concerning the WDW 
equation, the third quantization is investigated\cite{Faizal1, O-E1, O-E2} which is similar 
to the quantum field theory of the metric, if both theories were true ones, 
since both theories can treat the creation and annihilation of the universe.
However, the confrontation with the observation is very difficult.
Therefore we should first establish a reliable theory and then seek the 
confrontation of the theory with observations.
\\

\end{document}